\documentstyle[preprint,floats,tighten,epsf,prd,aps]{revtex}

\def\be{\begin{equation}}
\def\ee{\end{equation}}
\def\half{{1\over 2}}
\def\bea{\begin{eqnarray}}
\def\eea{\end{eqnarray}}
\def\bml{\begin{mathletters}}
\def\blea{\begin{mathletters}\begin{eqnarray}}
\def\elea{\end{eqnarray}\end{mathletters}}

\def\ba{{\bf a}}
\def\bb{{\bf b}}
\def\bx{{\bf x}}

\def\xdot{{\dot\bx}}
\def\sm{\sigma_-}
\def\sp{\sigma_+}
\def\xu{{\hat {\bf x}}}
\def\yu{{\hat {\bf y}}}
\def\zu{{\hat {\bf z}}}

\begin{document}
\draft
\title{Electromagnetic radiation from superconducting string cusps}

\author{J.\ J.\ Blanco-Pillado\footnote{Email address: {\tt
jose@cosmos2.phy.tufts.edu}} and  Ken D.\ Olum\footnote{Email address:
 {\tt kdo@alum.mit.edu}}}

\address{Institute of Cosmology, 
Department of Physics and Astronomy, 
Tufts University, 
Medford, Massachusetts 02155}

\date{August 2000}

\maketitle

\begin{abstract}%
Cusps in superconducting cosmic strings produce strongly beamed
electromagnetic radiation.  To calculate the energy emitted requires
taking into account the effect of the charge carriers on the string
motion, which has previously been done only heuristically.  Here, we
use the known exact solution to the equations of motion for the case
where the current is chiral to update previous calculations for the
total energy, spectrum and angular distribution in that case.  We
analyze the dependence of the radiated energy on the cusp parameters,
and discuss which types of cusp dominate the total radiation emitted
from an ensemble.
\end{abstract}


\narrowtext

\section{Introduction}
\begin{figure}
\begin{center}
\epsfxsize=4in
\leavevmode\epsfbox{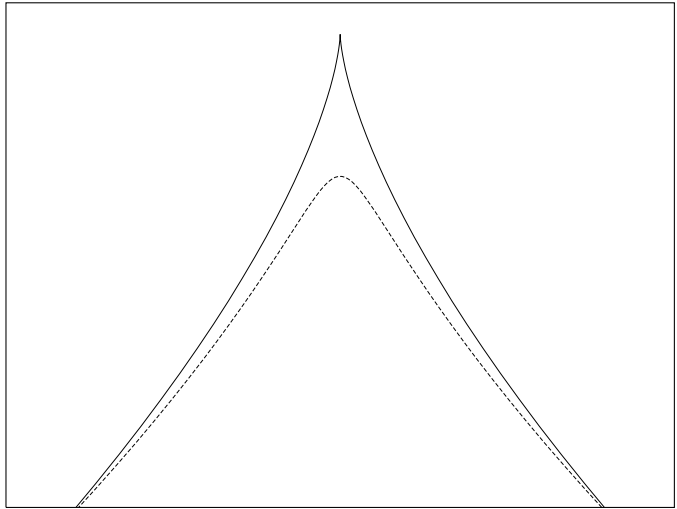}
\end{center}
\caption{Form of the cusp region for ordinary and 
superconducting cosmic strings: pure Nambu-Goto cusp (solid); including
charge carrier back-reaction (dashed).}
\label{fig:pseudo-cusp}
\end{figure}

Cosmic strings are one dimensional topological defects which may
have been left behind by phase transitions in the early
universe\cite{Kibble76,Alexbook}.  In the zero width approximation we
can describe the dynamics of the strings by the Nambu-Goto (NG)
action \cite{Nambu70,Goto71}, namely
\be
S = -\mu\int d^2\zeta\sqrt {- \gamma}
\ee
where $\mu$ is the energy per unit length of the string and $\gamma$ the
determinant of the worldsheet metric parameterized by the coordinates 
$\zeta^0$ and $\zeta^ 1$. Due to the reparameterization invariance
 of the Nambu-Goto action, we have the freedom to impose several conditions 
on the functions that denote the string position, in other words 
to fix the gauge. The most convenient gauge for ordinary strings
 in flat spacetime is the conformal gauge, in which we impose
\blea
|\bx' (\sigma, \tau)| ^ 2 + |\xdot (\sigma, \tau)|^2 & = & 1\label{eqn:xpxd}\\
\bx' (\sigma, \tau)\cdot\dot\bx (\sigma, \tau) & = & 0
\elea
where we have fixed $\zeta^0 = \tau = x^0$ and  $\zeta^1 = \sigma$,
and primes and dots denote differentiation with respect to $\sigma$ and
$\tau$ respectively. . With these conditions
the equations of motion become
\be
\bx'' (\sigma, \tau) =\ddot\bx (\sigma, \tau)\, .
\ee
The general solution is
\be\label{eqn:general}
\bx (\sigma, \tau) =\half (\ba (\sigma - \tau) +\bb (\sigma + \tau))
\ee
where $\ba$ and $\bb$ are arbitrary functions constrained by the gauge
 conditions to satisfy $|\ba' | =|\bb' | = 1$. 
From Eq.\ (\ref{eqn:general}) we see that
\be
\bx' =\half [\ba' (\sigma -\tau) + \bb' (\sigma +\tau)]
\ee
and
\be
\dot\bx =\half[ - \ba' (\sigma -\tau) + \bb' (\sigma +\tau)]\,.  
\ee
If there is a point along the string at which $\ba' = -\bb'$, then
$\bx' = 0$ and even more importantly $|\dot\bx | = 1$. This means that
there are points on the string that move at the speed of light and
where the string doubles back on itself. (See Fig.\
\ref{fig:pseudo-cusp}.) These events are called cusps\cite{Turok84}.

 It is clear that this kind of dynamics falls outside of the domain of
 validity of the assumptions involved to 
obtain the Nambu-Goto action. Nevertheless it has been
shown by field theory simulations that the motion of the string
is quite accurately described by the NG solution up to the point of
interaction between the two branches of the string\cite{jjkdo98.1}.

The violent motion of the string at the cusps suggests the
possibility of a range of effects which could have prospects of 
detection
\cite{Vachaspati85,Vilenkin87,Branden87,Babul87,Paczynski88,Pijus89,Branden90,Berezinsky-00,Damour00}.

In 1985 Witten\cite{Witten85} showed that certain particle physics models could lead 
to the formation of strings with superconducting behavior. It was
soon realized that these models could in principle have a much richer 
phenomenology due to the inertia of the charge carriers and 
their coupling to the long-range electromagnetic field.

In particular we can think that if the current on the string is not
very large, then the string motion would not be affected, and calculate the
electromagnetic radiation from a Nambu-Goto prescribed motion for the
string.  However, the power emitted by radiation near the cusp in this
approximation is divergent due to the infinite Lorentz factor of the
NG cusp\cite{Vilenkin87,Spergel87}. This infinite result suggests that
electromagnetic back-reaction is crucial in order to understand the
superconducting string cusp.

The short-range effects of the electromagnetic field can be taken into
account by renormalization of the charge carrier inertia and
charge\cite{Witten85,Copeland88,Alexbook}.  The charge carrier inertia
term prevents the appearance of an infinite Lorentz factor, and thus
the calculation yields a finite result.  In principle, this result
could be further modified by back-reaction due to long-range fields
(i.e. radiation), but for generic values of the parameters we expect
this correction to be small \cite{Blanco-Pillado00-1}.  Further
approximating that the current is chiral, at least in the vicinity of
the cusp, we are able to do a proper calculation of the total power
radiated.

\section{Superconducting strings}

The equations of motion for a superconducting string can be 
written\cite{Witten85,Vilenkin87}
\blea
\partial_a \left( \sqrt{- \gamma} \, \left(\mu \gamma^{ab} +
 \theta^{ab}\right) x^{\nu}_{, b} \right) &=& -\, \sqrt{- \gamma}\, F^{\nu}_{\sigma} 
x^{\sigma}_{,a} J^{a}\\
\partial_a\left(\sqrt{-\gamma}\gamma^{ab}\phi_{,b}\right) &=& - {1\over 2} q 
\epsilon^{ab} F_{\mu \nu} x^{\mu}_{, a} x^{\nu}_{, b} \\
\partial_{\nu}F^{\mu \nu} &=& - 4 \pi j^{\mu}
\elea
where $\mu$ is the energy per unit length of the string, $q$ the
electromagnetic coupling of the fields on the string,
$\theta^{ab}$ the energy-momentum tensor of the charge
carriers,
\be\label{eqn:fullthetadef}
\theta^{ab} = \gamma^{ac}\gamma^{bd}\phi_{,c} \, \phi_{,d} - {1\over
2} \, \gamma^{ab}\gamma^{cd} \, \phi_{,c} \, \phi_{,d}\, \,  ,
\ee
 $\phi$ the auxiliary scalar field in terms of which we can express
the conserved worldsheet current
\be\label{eqn:current}
J^a = q {1 \over{\sqrt{-\gamma}}} \,\epsilon^{ab} \, \phi_{,b} \, \, ,
\ee
and $j^{\mu}$ the 4-dimensional current,
\be\label{eqn:4D-current}
j^{\mu}(x) = \int{d^2 \zeta \sqrt{- \gamma}\,  J^a \, x^{\mu}_{, a}\, 
 \delta^{(4)}(x- x(\zeta))} \, .
\ee
These equations of motion are obtained from a two-dimensional effective
action for the superconducting string and they are valid as long the
current, measured in an invariant way, is not too large,
\be
J_a J^a \ll |J_{\rm crit}|^2 \,,
\ee
where $J_{\rm crit}$ is the limiting current at which charge carriers
can be ejected.
 
The leading effect of electromagnetic self-interaction of the
current-carrying string can be taken into account by a redefinition of
the auxiliary field $\phi$ and its
charge\cite{Witten85,Copeland88,Alexbook}. This enables us to include
the near-field effect on the left-hand side of the equations of
motion.  If we take this as our only back-reaction effect, then we can
rewrite the equations above in a much simpler way by taking $q$ and
$\phi$ to be the renormalized quantities, to get
\blea
\partial_a \left( \sqrt{- \gamma} \, \left(\mu \gamma^{ab} +
 {\theta}^{ab}\right) x^{\nu}_{, b} \right) &=& 0\\
\partial_a\left(\sqrt{-\gamma}\gamma^{ab} {\phi}_{,b}\right) &=& 0 \\
\label{eqn:renorm-equation-3}
\partial_{\nu}F^{\mu \nu} &=& - 4 \pi j^{\mu}
\elea

This procedure decouples the first two equations from the third one, 
and they now give the motion of a `neutral' superconducting string.
In the last equation, $F^{\mu\nu}$ is the external field emitted by
the current moving according to the first two equations.

 Numerical simulations\cite{Davis88-2,Martins98} have suggested that
as a superconducting cosmic string loop contracts under its own
tension, it would approach the `chiral' limit, in which the current on
the string is null, in other words, $J_a J^a = 0$. This, in turn,
means that the auxiliary scalar field fulfills the condition,
\be\label{eqn:chirality} 
\gamma^{ab}\phi_{, b}\phi_{, a} = 0 \, .
\ee

Furthermore, it can be shown\cite{Copeland88} that if the current
is not chiral it would become supercritical around the cusp and charge
carriers would be ejected.  We therefore expect that this process
would drive this region of string around the cusp to the chiral
limit. These arguments suggest that most of the electromagnetic
radiation would come from chiral current carrying string cusps.
 
If the current is chiral, the set of equations describing the string and the 
`neutral' current can be solved completely\cite{Carter99-1,Blanco-Pillado00-1,Davis00-1}.
In this case, the general solution for the scalar field is
\be
\phi (\sigma,\tau) = F (\sigma +\tau)\,,
\ee
where $F$ is an arbitrary function, and the string position is given by
\blea
x^0 & = & \tau \\
\bx & = & {1\over 2}[\ba(\sigma - \tau) + \bb(\sigma + \tau)]\,, 
\elea
with constraints for the otherwise arbitrary functions $\ba'$ and $\bb'$,

\bml
\label{eqn:constraints}
\bea
|\ba'|^2 &=& 1 \\
|\bb'|^2 &=& 1 -  {{4 \, F'^2}\over {\mu}}\,.
\elea

This description of the chiral strings makes evident that they behave
very similarly to the Nambu-Goto case. As before, the functions $\ba'$
and $\bb'$ live on the surface of spheres (if the current is
constant), but now their radii are different for the two cases.  This
change, however, brings important consequences for the string
motion. As the magnitude of the current increases, the average
velocity of the loop decreases, arriving at the limiting stationary
case when $4\, {F'}^2 /\mu = 1$, the chiral vorton\cite{Davis88-2}.

This also means that the string can only reach a finite
Lorentz factor at the cusp-like regions, in contrast with the Nambu-Goto case.
This effect can be easily understood in terms of the resistance of the charge
carriers to being accelerated by the string tension.  Equations (\ref{eqn:constraints}) also imply that $|\bx'|>0$ at the cusp so the 
actual shape of the string around the cusp is rounded off by the effect of the
charge carriers, as shown in Fig.\ \ref{fig:pseudo-cusp}.  (Another
way for the string to be rounded off, producing a self intersection,
is discussed in \cite{jjkdo00-3}.)

\section{Electromagnetic Radiation from the cusp}

Using the solution for the string motion from the previous section,
which as discussed is free from divergences in its Lorentz factor, we
can now compute, using Eq.\ (\ref{eqn:renorm-equation-3}), the
electromagnetic radiation from the cusp-like region. Throughout this
section we follow closely the calculation of the radiation by Spergel
{\it et al.}\cite{Spergel87}, and introduce the necessary modifications
for the chiral string motion.

Since the radiation is mainly produced by the region with a 
large Lorentz factor, we can estimate the energy output from the position 
of the string expanded in a Taylor series around the cusp, which we take as 
$\sigma=0$ and $\tau = 0$. We also assume throughout the calculation that 
${2 \, |F'|/ {\sqrt{\mu}}} \ll 1$, since otherwise the string would not reach
 high Lorentz factors at all. 

Using the light-like variables $\sigma_+ = \sigma + \tau$ and $\sigma_- = \sigma - \tau$,
we can expand the functions $\ba$ and $\bb$ near $\sigma_{\pm}=0$
as
\blea
\ba(\sm) & = & \ba'_0 \sm + {1\over 2} \ba''_0 \sm^{2} + {1\over 6} \ba'''_0 \sm^{3} + \cdots \\
\bb(\sp) & = & \bb'_0 \sp + {1\over 2} \bb''_0 \sp^{2} + {1\over 6} \bb'''_0 \sp^{3} + \cdots \, ,
\elea
where we center our coordinate system at the position of the cusp, so
that $\bx_0 = 0$.  In analogy with the Nambu-Goto case, we consider
the cusp to be the place at which $\ba'$ and $\bb'$ are anti-parallel,

\be
{{\ba'_0}\over {|\ba'_0|}} = - {{\bb'_0}\over {|\bb'_0|}}
\ee
which corresponds to the point of maximum concentration of energy and the minimum value
of $|\bx'|$\cite{Blanco-Pillado00-1}.
Assuming that the current remains constant over the region of interest, we can see that the
constraint equations (\ref{eqn:constraints}) impose a constant magnitude for the vectors $\ba'$ and $\bb'$. 
We can choose for simplicity 
\bml\label{eqn:a0b0}\bea 
\ba'_0 &=& - \xu \, \\
\bb'_0 &=& \Delta \xu 
\elea
where $\Delta = 1 - \epsilon$, with $\epsilon \ll 1$ and
constant.  Due to the fact that
$|\ba'|$ and $|\bb'|$ are constants we have the following
relations,
\bml\label{eqn:a0b0-3}\bea
\ba'_0 \cdot \ba''_0 &=& \bb'_0 \cdot \bb''_0 = 0 \, \\
\ba'_0 \cdot \ba'''_0 &=& - {|\ba''_0|}^2 \, \\
\bb'_0 \cdot \bb'''_0 &=& - {|\bb''_0|}^2
\elea

Using Eqs.\ (\ref{eqn:a0b0}, \ref{eqn:a0b0-3}) we can express the most general 
expansion of the string position near the cusp up to third order in
 $\sigma_{\pm}$ in the form
\blea
\ba(\sm) & = & \left({1\over 6}\, \alpha^2 \, \sm^3 - \sm\right) \, \xu 
+ \left({1\over 2}\, \alpha \, \sm^{2} + {1\over 6} \,\varrho \,\sm^3\right) \, \yu + \left({1\over 6}\, \delta \,\sm^{3}\right) \, 
\zu \\
\bb(\sp) & = & \left( \Delta \, \sp - {{(\beta^2 + \gamma^2)}\over {6 \Delta}} \, \sp^3\right) \, \xu 
+ \left({1\over 2}\, \beta \, \sp^{2} + {1\over 6} \,\varpi \,\sp^3\right) \, \yu + \left({1\over2} \gamma \sp^2 + {1\over 6}\, \eta \,\sp^{3}\right) \, \zu\,. 
\elea
The parameters $\alpha, \beta, \gamma$ are of order $L^{-1}$, with $L$ denoting the 
typical length scale of the loop, and $\delta, \varpi, \varrho, \eta$
are of order $L^{-2}$.

Using standard results in electromagnetic theory\cite{Jackson}, we can
calculate the energy radiated per unit frequency and per solid angle
for a source given by Eq.\ (\ref{eqn:4D-current}) using the
expression\cite{Spergel87},
\be
{{d^2 E_{rad}}\over {d\omega \, d\Omega}} = {{{\omega}^2}\over {4 {\pi}^2}} 
\left|\int{ d\sigma \, \int{ d\tau ({\bf n} \times {\bf c})\, \exp[ - i \omega 
(n^{\mu} x_{\mu})]}}\right|^2
\ee
where $n^{\mu}$ is the unit null vector from the source point to the observation
point,

\be
n^{\mu} = \bigg{(}1 , \cos \theta, \sin\theta \, \cos\varphi , \sin\theta \, \sin\varphi\bigg{)} \simeq 
 \bigg{(}1 , 1 - {{\theta^2} \over 2}, \theta \, \cos\varphi , \theta \, \sin\varphi\bigg{)} \, .
\ee
This last approximation is justified since the radiation is highly beamed in the direction of
movement of the cusp, in this case, ${\bf \hat{x}}$. The four-vector
$c^{\mu}$ is the source current, which in our case takes the form
\be
c^{\mu} = q \,  \epsilon^{ab} \, \partial_a \phi \, \partial_b x^{\mu}
\ee
where $a,b =1,2$  stand for $\sigma$ and $\tau$.
In the chiral case, the four-vector current becomes
\be
c^{\mu} = q \, {{d\phi}\over {d\sigma}} \, \left({{\partial x^{\mu}}\over {\partial \sigma}} - 
{{\partial x^{\mu}}\over {\partial \tau}}\right) = 2 q F' \, {{\partial x^{\mu}}\over {\partial \sigma_{-}}} \, .
\ee
Keeping only the lowest terms in the angle $\theta$, we get
\bea\label{ded2}
{{d^2 E_{rad}}\over {d\omega \, d\Omega}} & = &  {{{\omega}^2}\over {16 \pi^2}} 
\, \left|\int{d\sigma_- \, \, {\bf {\cal C}}(\sigma_-,\theta,\varphi,F') \, \, 
\exp[ i \, \omega \, (J \sm + K \sm^2 + L \sm^3)]}\right|^2 \nonumber \\
&& \times \left|\int{ d\sigma_+ \, \exp[- i\,  \omega \, (S \sp + T \sp^2 + U \sp^3)]}\right|^2
\eea
with

\be
{\bf {\cal C}}(\sm,\theta,\varphi,F') =  q F' \bigg{(}- \alpha \theta \sin \varphi \, \sm, - \theta \sin \varphi,
 \alpha \sm + \theta \cos \varphi \bigg{)}
\ee
and 

\blea
J &=&  {{\theta^2}\over 4} \\
K &=&  {{\theta}\over 4} \alpha \cos \varphi  \\
L &=&  {{\alpha^2}\over 12}
\elea
and 

\blea
S &=& {1\over 2} \left(1 - \Delta (1 - {{\theta^2}\over 2})\right) \\
T &=& - {{\theta}\over 4}   \left(\beta \cos \varphi + \gamma \sin \varphi\right) \\
U &=&  {{1}\over {12 \Delta}} \left(\beta^2 + \gamma^2 \right) \, .
\elea

The two integrals in Eq.\ (\ref{ded2}) can be evaluated, first
changing variables to get rid of the term quadratic in $\sigma_{\pm}$
in the exponent, and then using the following forms of the Airy integrals:
\be 
 \int_{0}^{\infty}{dx \cos\left[{3\over 2} \xi \left(x + {1\over 3} x^3\right) \right]} =
{1\over {\sqrt{3}}} \,  K_{1\over 3} (\xi)
\ee
and
\be 
 \int_{0}^{\infty}{dx \, x \, \sin\left[{3\over 2} \xi \left(x + {1\over 3} x^3\right) \right]} =
{1\over {\sqrt{3}}} \,  K_{2\over 3} (\xi)
\ee
where $K_{\nu}$ are the modified Bessel functions of order $\nu$.
With a little bit of algebra we get to the result,
\bea\label{eqn:dedwdo}
{{d^2 E_{rad}}\over {d\omega \, d\Omega}} & = & {{ {\omega}^2}\over {16 \pi^2}}
\, {\left(q F' \right)}^2 \, \left({2\over 3} \, K_{1\over 3}(\xi_+) \, \chi_+  \right)^2 
  \nonumber \\
&& \times 
 \left[ \theta^2 \, \sin^2 \varphi  
\left({2\over 3} \, K_{1\over 3}(\xi_-) \chi_- \right)^2
 + \alpha^2 \, \left({2\over {3  \sqrt{3}}} \, K_{2\over 3}(\xi_-) \, \chi_-^2 \right)^2 \right]
\eea
where
\blea
\chi_{+}(\epsilon) & = & \sqrt{{3SU - T^2}\over {3U^2}} = 
\left({{3\, [2\, \epsilon +
 \theta^2 \, Y]}\over {\beta^2 + \gamma^2}}\right)^{1/2}\\
\label{eqn:xi+}
\xi_{+}(\epsilon) & = & {2\over {3\sqrt{3}}} \omega
 \left(S - {{T^2}\over {3U}}\right) \,
 \chi_{+}= {1\over 6} \omega\left({{ \left[2\, \epsilon +
 \theta^2 \, Y\right]^3}\over \beta^2 + \gamma^2}\right)^{1/2}
\elea 
and where we have defined the dimensionless quantity,
\be
Y = {{\left(\beta \sin \varphi -\gamma \cos \varphi\right)^2}\over{\beta^2 + \gamma^2 }}= \sin^2 (\varphi - \varphi_0) \,, 
\ee
where $\varphi_0$ is the angle between $\ba''_0$ and $\bb''_0$, so that
$\tan \varphi_0= \gamma/\beta$. 
On the other hand the expression for the same type of parameters for the
integral in $\sm$ are much simpler,
\blea
\chi_{-} & = & \sqrt{{3JL - K^2}\over {3L^2}} = \sqrt{3} \, \theta \,\left|
{{\sin \varphi}\over {\alpha}} \right|\\
\xi_{-} & = & {2\over {3\sqrt{3}}} \, \omega \, \left(J - {{K^2}\over {3L}}\right) \, \chi_{-} = {1\over 6} \,
\omega \, \theta^{3} \, \left|{{\sin^3 \varphi}\over { \alpha}}\right| \, .
\elea 
It is important to note that the different behavior of the superconducting
chiral string cusp is encoded in the presence of the parameter $\epsilon$.
If we take the limit $\epsilon = 0$, we would recover the pure
Nambu-Goto cusp. 

Using the above  definitions, we can simplify Eq.\ (\ref{eqn:dedwdo}) to

\be
{{d^2 E_{rad}}\over {d\omega \, d\Omega}}  =  \left({{q F' \sin^2 \varphi}\over{3 \pi \alpha}}\right)^2 \,
\left( \omega \, \theta^2\right)^2 \, 
\left({{ 2\, \epsilon +
 \theta^2 \, Y}\over {\beta^2 + \gamma^2}}\right) 
 \left[K_{1\over 3}(\xi_+(\epsilon))\right]^2 \,
\left( \left[K_{1\over 3}(\xi_-)\right]^2
 + \left[K_{2\over 3}(\xi_-)\right]^2 \right)
\ee
We can now integrate over frequencies with the change of variables $z
=\xi_-$ to get,
\be\label{eqn:dedwdo-2}
{{d E_{rad}}\over {d\Omega}}  = 24 \, \left({q F'\over \pi}\right)^2 \,
 \theta^{-3} \, |\sin^{-3} \varphi| \,
{|\alpha|^{1/3} \over\left(\beta^2 + \gamma^2\right)^{2/3} }\,
{\cal F} ({\xi_+ / \xi_-})
\ee
where we have defined the function ${\cal F}$ as
\be 
{\cal F}(a) =
\int_{0}^{\infty} dz \,a^{2/3}  z^2 \,
\left[\left[K_{1\over 3} (a z) \, 
K_{1\over 3} (z) \right]^2 +
\left[K_{1\over 3} (a z) \, 
K_{2\over 3} (z) \right]^2\right] \, .
\ee 
The integral can be done exactly in hypergeometric functions.  The
function ${\cal F}$ has asymptotic behavior
\be
{\cal F} (a)\to\cases{\rm const & $a\to 0$\cr {\rm const}/a & $a\to\infty$}
\ee
(with different constants).
\pagebreak[1]

\begin{figure}
\begin{center}
\epsfxsize=4in
\begin{picture}(300,200)
\put(-40,120){{\Large{${\cal A}$}/L}}
\put(140,5){{\Large{$\varphi_0$}}}
\put(12,28){$0$}
\put(82,28){$\pi/4$}
\put(144,28){$\pi/2$}
\put(210,28){$3\pi/4$}
\put(285,28){$\pi$}
\put(1,36){\leavevmode\epsfbox{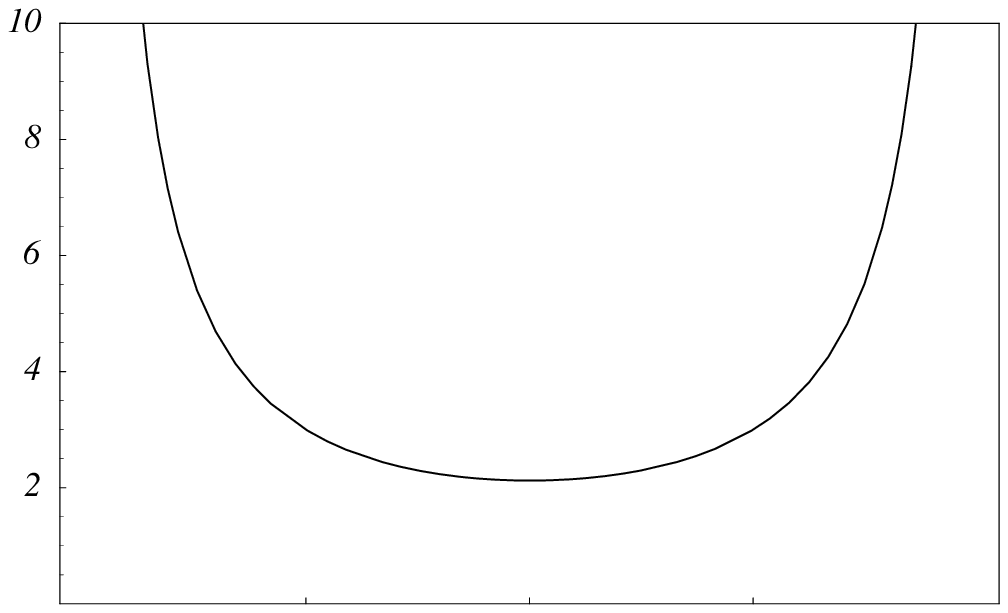}}
\end{picture}
\end{center}
\caption{Numerical coefficient ${\cal A}$ for an $m = 1$, $n = 2$ Burden loop.}\label{fig:A-Plot}
\end{figure}

In order to integrate Eq.\ (\ref{eqn:dedwdo-2}) with respect to the angle 
$\theta$, we first note that we can extend the range of integration to
infinity since only the region around the direction of the cusp contributes.
In order to extract the dependence on the parameter $\epsilon$, we perform
 another change of variables to $\theta'=\theta/\sqrt{\epsilon}$, so the
final expression for the energy emitted at the cusp becomes
\be\label{eqn:totalE}
E_{rad} = {\cal A}(\alpha,\beta,\gamma) \, (q F')^2 \, \epsilon^{-1/2}\,,
\ee
where ${\cal A}(\alpha,\beta,\gamma)$ is a constant which can be written in
 terms of the magnitudes of $\ba''_0$ and $\bb''_0$ and the angle between 
them, $\varphi_0$, as
\be
{\cal A} =  {24\over \pi^2}
{|\ba''_0|^{1/3}\over {|\bb''_0|^{4/3}}}  \int_0^{2\pi} d\varphi
\int_0^{\infty} d\theta'\, 
\theta'^{-2} |\sin^{-3}\varphi|
 {\cal F} \left[{|\ba''_0| \over |\bb''_0|} 
 \,\left({{2 + \theta'^2 \, \sin^2 (\varphi -\varphi_0)}
\over {\theta'^2  \sin^2 \varphi}}\right)^{3/2}\right]
\sim O(L) \,.
\ee
Figure \ref{fig:A-Plot} plots the value of ${\cal A}/L$ versus $\varphi_0$ 
for an $m = 1$, $n = 2$ Burden\cite{Burden:1985md} loop, which has
 $|\ba''_0|= 2\pi/L$ and $|\bb''_0|=4\pi/L$ at the cusp.

We can understand the form of ${\cal A}$ as follows.
When $\theta'\ll 1$ (i.e., $\theta\ll\sqrt{\epsilon}$), the argument
to ${\cal F}$ goes as $\theta'^{-3}\sin^{-3}\varphi\gg1$, and the
integrand becomes proportional to $\theta'$, which is just the
geometrical factor from the integration.  Thus
the radiated power per solid angle is uniform for small angles.  Now
we consider the case with $\theta'$ large.  When $\sin(\varphi
-\varphi_0)$ is far from zero, then $\theta'\gg1$ makes the argument
of ${\cal F}$ constant.  In such directions, the radiated power is
suppressed by $\theta'^{-2}$, so their contribution is small.  If
$\sin(\varphi -\varphi_0)$ is small, then the argument of ${\cal F}$
goes to zero, and thus the value to a constant, unless $\sin\varphi$
is also small.  Thus, as long as $\varphi_0 $ is far from $0$ and $\pi$,
the radiation is confined to a beam with radius
$\theta\sim\sqrt{\epsilon}$.

However, if $\varphi_0$ is near $0$ (or, equivalently, near $\pi$),
then it is possible to have $\sin (\varphi -\varphi_0)$ and
$\sin\varphi$ small simultaneously. In that case, there is a range of
$\theta'$ up to about $\varphi_0^{-1}$, in which large contributions to
the integral are still possible.  This is the cause of the divergence
seen in Fig.\ \ref{fig:A-Plot}: ${\cal A}$ is proportional to
$\varphi_0^{-1}$ for $\varphi_0\ll1$.  In this case, the beam can be
greatly elongated in the plane of the cusp.  In the limiting case,
$\phi = 0$ or $\phi =\pi$, the Burden loop has a cusp at all times,
which rotates around the string.  In cases close to this, we see that
the cusp is longer lived than usual and thus has the opportunity to
beam radiation through a wider angle.

In this regime, Eq.\ (\ref{eqn:totalE}) is not correct, because we
cannot extend the $\theta$ integration to infinity.
Instead, we should have a cutoff at $\theta'\sim\epsilon^{-1/2}$.  If
we include this effect, then the growth of ${\cal A}$ with
$\varphi_0^{-1}$ is cut off at $\varphi_0\sim\sqrt{\epsilon}$.

Now suppose that we have a set of loops with $\varphi_0$ distributed
evenly between $0$ and $\pi$.  To find the contribution due to the
loops with $\varphi_0$ near 0, we can integrate
\be
\int_0d\varphi_0\,E (\varphi_0)\sim 
\int_{\epsilon^{-1/2}} d\phi_0\,\phi_0^{-1}\sim-\ln\epsilon \,.
\ee 
Thus if $\epsilon$ is small, loops with $\ba''_0$ and $\bb''_0$ nearly
parallel or antiparallel dominate the total electromagnetic radiation,
but the effect is only logarithmic.

Returning to the generic case, let's now remember that we have defined
$\epsilon$ to be the deviation of the magnitude $\bb'$ from unity,

\be
|\bb'| = \Delta = 1 - \epsilon
\ee
so in the chiral string model,  for a current much smaller than
the energy per unit length of the string,
\be
|\bb'| = \sqrt{1 - {{4 |F'|^2}\over {\mu}}} \approx 
1 - {{4 |F'|^2}\over {2\mu}}
\ee
so
\be
\epsilon \approx {{4 |F'|^2}\over {2\mu}}
\ee

It can be shown\cite{Blanco-Pillado00-1} that the parameter $\epsilon$ 
determines the maximum Lorentz factor reached near the cusp, namely,
\be
\Gamma_{max} = (2 \epsilon)^{-1/2}
\ee
so we can rewrite Eq (\ref{eqn:totalE}) as
\be\label{eqn:result1}
E_{rad} \sim q^2 \,\sqrt{\mu}\, |F'| \, L \sim  L \,  (q F')^2 \, \Gamma_{max}
\ee
in agreement with earlier estimates\cite{Vilenkin87,Spergel87}. We can also
write the total energy output in terms of the physical current as
\be\label{eqn:result2}
E_{rad} \sim q \, j \,\sqrt{\mu}\, L \, .
\ee
Equations (\ref{eqn:result1},\ref{eqn:result2}) are valid when
$\phi_0$ is far from $0$ and $\pi$.

\section{Discussion}
Electromagnetic radiation from superconducting cosmic strings cusps
has traditionally been considered a distinctive signature of
superconducting string models. Recently this interest has been revived
by as possible connection with gamma ray
bursts\cite{Berezinsky-00,Damour00} and ultra-high energy cosmic rays.
The total energy output from a pure Nambu-Goto string cusp (i.e. no
back-reaction included) yields an infinite result due to the infinite
Lorentz factor of the tip of the string, where the charge carriers get
concentrated. It is therefore necessary to impose some sort of cutoff
for this process. Early calculations\cite{Vilenkin87,Spergel87}
assumed the cutoff to be where the energy in the charge carriers was
comparable with the energy of the string. This procedure gives a
linear dependence of the total energy with the maximum Lorentz
factor. We redo this calculation in the context of a chiral current on
the string. This is the most interesting case since loops are driven
to chirality by charge carrier ejection\cite{Davis88-2,Martins98},
especially near the cusps\cite{Blanco-Pillado00-1}.

The full system of equations for a superconducting string is very
complicated. Nevertheless in the case of a chiral current they are
much simpler and can be solved
exactly\cite{Carter99-1,Blanco-Pillado00-1,Davis00-1}. This allows us
to compute the electromagnetic radiation from a chiral-current string
trajectory taking into account in an exact way the back-reaction of the
charge carriers on the string motion. In order to do this we closely
follow the procedure described in Spergel {\it et al.}\cite{Spergel87}
and we extract the dependence of the total energy on the maximum
Lorentz factor that the string has in  the pseudo-cusp region. We found
the same linear dependence in the maximum Lorentz factor as earlier
estimates.  This suggests that the intuitive picture of thinking about
the back-reaction of the charge carriers as a process that only becomes
important when their energy density is comparable to the energy of the
string is correct, making it possible to extend this result to all types
of currents.

In the case of cusps where the parameters $\ba''_0 $ and $\bb''_0 $ are
nearly parallel or antiparallel, the radiation can be much larger than
in the generic case.  For small currents, the radiation from such
cusps dominates in the total flux, but only by a logarithmic factor.

\section{Acknowledgments}

We would like to thank Alex Vilenkin and Xavier Siemens for helpful
conversations. This work was supported in part by funding 
provided by the National Science Foundation.


\end{document}